\documentclass[preprint]{aastex}

\shortauthors{Palmer \& Goss \& Devine}
\shorttitle{VLBA Observations of OH Masers}

\slugcomment{Aug 15, 2003}
\begin{document}

\title{Phase-Referenced VLBA Observations of OH Masers at 4765 MHz}

\author{Patrick Palmer}
\affil{Department of Astronomy and Astrophysics, University of Chicago,
5640 S. Ellis Ave., Chicago, IL 60637}
\email{ppalmer@oskar.uchicago.edu}
\author{W. M. Goss}
\affil {National Radio Astronomy Observatory, Socorro, NM 87801}
\email{mgoss@nrao.edu}
\and 
\author{K. E. Devine\footnote{current address:  University of Wisconsin,
Department of Astronomy, 475 N. Charter Street, Madison, WI 53706}}
\affil{Carleton College, Northfield, MN 55057}
\email{devinek@carleton.edu}

\begin{abstract}

We report VLBA observations of maser emission from the rotationally excited
$^2\Pi_{1/2}, J=1/2$ state of OH at 4765 MHz.  We made phase-referenced
observations of W3(OH) at both 4765 MHz and 1720 MHz and found emission in
three fields within a $\sim$2000 AU diameter region and verified that in
two of the three fields, 4765 MHz and 1720 MHz emission arise from the same
position to within $\sim$4 mas ($\sim$8 AU).  We imaged DR21EX without
phase-referencing and detected six $\sim$5 AU diameter emission regions
along an approximately N-S arc with linear extent $\sim$500 AU.  In
addition, we carried out phase-referenced observations of 4765 MHz emission
from K3-50.  We searched for the 4765 MHz line in W49 (without phase
referencing) and W75N (phase-referenced to the strongest 4765 MHz maser
feature in DR21EX); we were unable to detect these sources with the
VLBA.  For 2 1/2 years (including the dates of the VLBA observations), we
carried out monitoring observations of 4765 MHz emission with the VLA.
Constraints on models for maser emission at 1720 MHz and 4765 MHz are
derived from the observations.  These observations are then briefly
compared with existing models.

\end{abstract}


\section{Introduction}
OH masers have long been associated with the early stages of star
formation.  Models of maser pumping mechanisms allow constraints to be set
on the physical conditions in the region of the maser.  By understanding
the conditions associated with OH masers, the conditions in regions of star
formation can be delineated.  Many models have been developed for pumping
mechanisms for the ground state (18 cm wavelength) lines; a number of the
models have been extended to excited state masers.  One of these excited
states is the $^2\Pi_{1/2}, J=1/2$.  The most commonly detected line from
this state is the satellite F=$1\rightarrow 0$ transition at 4765 MHz.
Although there are several models for emission from this state
(e.g. \citealt{kn90}; \citealt{cw91}; \citealt{gray92}; and
\citealt{pk96a,pk96b}), these models are not
entirely satisfactory.  Detailed observations are required including
higher spatial resolution and absolute position determinations.

Previous studies with increasing accuracies have shown many instances of
spatial coincidence between 1720 MHz ground state masers and 4765 MHz
masers (\citealt{cw91}; \citealt{mfg94}; \citealt{gray01} (hereafter
GCRYF)).  The possible spatial overlap has been an important factor in
recent pumping models.  In particular, the study of GCRYF with MERLIN
provided the first absolute positions of 4765 MHz and 1720 MHz masers
in W3(OH).  To within the uncertainties (estimated to be 5 mas at 4765 MHz
and 15 mas at 1720 MHz), they found that in two of three fields the
positions of the two masers agreed.  In addition, they compared angular
sizes of 1720 MHz masers \citep{mfg94} to those of accompanying 4765 MHz
masers and found that the 4765 MHz masers are considerably larger, implying
a smaller gain at the higher frequency.

Another important feature of 4765 MHz masers is their polarization
properties.  Theoretically, if polarization of molecular emission lines
arises from magnetic effects, these lines should be unpolarized because the
Land\'{e} g-factor of this state is $\sim$0 \citep{dst55}.  Probably
because because polarization is not expected, little observational effort
has been expended to measure polarization of these lines.  To date
polarization has been detected in only one source.  During a spectacular
flare, the 4765 MHz maser in MonR2 was found to be 15\% linearly polarized
\citep{sch98}.  Amplification of an unseen polarized background source was
a suggested cause.  Recently, \citet{de02} obtained upper limits for
polarization in a survey of southern star-forming regions.  The most
stringent limits for the 16 sources in which 4765 MHz emission was detected
were $\leq$4\% linear and $\leq$5\% circular.

Spatial coincidence with maser emission in other OH transitions, angular
size measurements, and polarization all provide information about possible
pumping mechanisms in these masers.  Furthermore, existing models for 4765
MHz emission focus largely on W3(OH). By observing a number of other
sources, we are able to examine whether these models are
more generally applicable.

In this paper we report VLBA observations of 4765 MHz masers in W3(OH),
DR21EX, and K3-50 as well as negative results for W49 and W75N.  We also
provide a preliminary report of VLBA observations of 1720 MHz emission in
W3(OH).

Depending on the availability of a nearby reference source, the VLBA
observations were made using phase-referencing.  Phase-referencing serves
two important purposes in this study: first, this technique provides a good
initial phase correction so that narrow, relatively weak maser lines (peak
flux densities $\leq$1 Jy) can be imaged with the VLBA; second, this
technique permits absolute position determinations.  This experiment
improves previous investigations for the following reasons: 1) the
synthesized beam of the VLBA is about an order of magnitude smaller that
the MERLIN beam, 2) we measured all four Stokes parameters for 4765 MHz
masers, and 3) we observed additional sources.

\section{VLBA Observations and Analysis} 
All 4765 MHz observations were carried out using the National Radio
Astronomy Observatory\footnote{The National Radio Astronomy Observatory is
a facility of the National Science Foundation operated under cooperative
agreement by Associated Universities, Inc.} (NRAO) Very Long Baseline Array
(VLBA) in 2000 December 23 (W3(OH)) and 2001 February 19 and 24 (K3-50,
DR21EX, W75N, and W49).  Observations of the latter sources were
interleaved over the two days to optimize hour angle coverage.
Phase-referencing on nearby calibrators was used for W3(OH) and K3-50:
scans of 3 minutes duration were made on a target source followed by 2
minute duration observations of the nearby reference source (see
Table~\ref{tab:tab1}).  Because of the current paucity of calibrators near
the Galactic plane, phase-referencing could not be used for W49, W75N, and
DR21EX.  However, because DR21EX has some very strong features, these could
be used as the phase-reference calibrator to determine phase corrections
for W75N (although, of course not for the absolute position).  For DR21EX
and W75N, 3 minute scans were made on each.  Ten antennas were used in all
observations: either all VLBA antennas or (in December) one antenna from
the Very Large Array (VLA) was substituted for the Pie Town antenna.  Data
was taken in 128 spectral channels spaced by 3.91 kHz in all four
polarization combinations.  With uniform weighting of the spectrum, this
resulted in a velocity resolution of 0.29 km s$^{-1}$ across $\sim$32 km
s$^{-1}$ from which all four Stokes parameters could be imaged.  When
phase-referencing was used, the accuracy of the absolute positions reported
here is dominated by residual effects which were not removed by
phase-referencing.  This error is estimated to be $\pm$1 mas at 6 cm.  This
error is dominant when comparing our results to those of others or our
results at 4765 MHz to those at 1720 MHz.  The relative errors for
positions at the same frequency in the same source are determined by
thermal noise and are $\sim$0.1 mas for all reported measurements.

The 1720 MHz observations of W3(OH) were carried out with the VLBA on 2002
July 31.  All 10 antennas of the VLBA were used.  Phase-referencing was
used with the same cycle as used for the 4765 MHz observations W3(OH).
Data was taken in 1024 spectral channels spaced by 0.488 kHz.  With uniform
weighting, this resulted in a velocity resolution of 0.085 km s$^{-1}$
across $\sim$80 km s$^{-1}$.  Because these lines were previously known to
be strongly circularly polarized, only right and left circular were
observed.  As at 4765 MHz, the accuracy of the absolute positions
determinations is dominated by residual ionospheric effects which were not
removed by phase-referencing.  This error is estimated to be $\pm$5 mas at
1720 MHz.  For all reported results, the relative position errors
determined by thermal noise are $\sim$0.1 mas.

Pointing positions were obtained from VLA measurements, and therefore are
accurate at the 1\arcsec \ level (\citealt{gwp1}; \citealt{pwg03}).  The
pointing positions\footnote{Most VLA positions quoted in this paper and the
positions input for the VLBA observations are in the B1950 system.  This
convention was for compatibility with earlier measurements.  In this paper,
positions obtained from VLBA observations will be given in the J2000
system, and we will use this system for future measurements.}, calibrators,
distances to the reference sources, and total times on source are provided
in Table~\ref{tab:tab1}.

\clearpage

\begin{table}
\caption{Pointing Positions and Calibrators}
\label{tab:tab1}
\begin{tabular}{lllccclc}
\tableline \tableline
Source & $\alpha_{1950}$ & $\delta_{1950}$ & Phase-Reference & Distance &
Bandpass & Total \\ 
        &             &    & Source & (degree) & Calibrator  & (minute) \\ 
\tableline
W3(OH) & &  &  &   &  &  \\
\ \ \ 4765 MHz:  &02 23 16.45&61 38 57.51 &0223+671 & 5.5 &DA193 & 540 \\
\ \ \ 1720 MHz: &02 23 16.40&61 38 57.30 &0223+671 & 5.5 & 3C454.3 & 460 \\
K3-50&19 59 50.172&33 24 21.42 & J2025+3343 & 4.9 &3C454.3 & 104 \\
DR21EX&20 37 13.552&42 14 01.23 & --- & --- & 3C454.3 & 106 \\
W75N&20 36 50.11&42 26 58.49 & DR21EX & 0.2 & 3C454.3 & 101 \\
W49&19 07 47.319&09 00 20.53 &---  & ---& 3C454.3 & 90 \\ \tableline
\end{tabular}
\end{table}

\clearpage

The data reduction was performed using standard AIPS software, following
the guidelines set out in Appendix C of the AIPS Cookbook.  This appendix
contains instructions for spectral line phase-referenced imaging.  During
the final calibration steps, SETJY and CVEL were added to the steps
outlined in the Cookbook before the data were split and imaged.  Self
calibration was used to improve the signal to noise ratio of the images;
the rms noise level obtained was about 5 mJy per beam.  During reduction,
data from the Saint Croix and Mauna Kea antennas were deleted because the
sources were resolved on these long baselines.  Images were then made of
all the spectral channels using the AIPS task IMAGR.  At 4765 MHz, images
were made for all four Stokes parameters; at 1720 MHz, images were made
separately for left and right circular polarization.  The synthesized beam
was typically 3 -- 4 mas at 4765 MHz and $\sim$6 mas at 1720 MHz for the
resulting images.  Detailed results for each source are provided in Section
4.

\section{VLA Monitoring Observations}

During the course of the experiment and shortly afterward, the sources
observed with the VLBA at 4765 MHz were monitored with the VLA.  Monitoring
was important so that any significant time variability could be identified,
and so that the flux densities observed with the VLBA could be compared
with those observed with the VLA to determine the fraction of flux density
resolved out with the VLBA.

The observations are summarized in Table 2. 
Both different configurations of the VLA and different spectrometer setups
were used, resulting in varying spatial (third column) and spectral (fifth
column) resolutions.  In all cases, the VLA data were Hanning smoothed.
The line emission was unresolved spatially, except for the A-array
observations of W3(OH) on 2002 January 11 and April 22.  In these
observations, three closely spaced point sources are obvious which
correspond to the three fields studied in more detail with the VLBA
(Section 4.1).  Because of the varying spectral resolution, we integrated
the flux density in each spectrum and computed velocity centroids; thus,
we could compare quantities that do not depend on spectral resolution.  The
observed velocity-integrated fluxes and the velocity centroids for W3(OH),
DR21EX, and K3-50 are provided in Table~\ref{tab:vlavlba}.  Although there
may be variability at the 10 -- 20 percent level, there is no evidence for
episodes of large variability in the 2 1/2 years of observations of
W3(OH), DR21EX, and K3-50 reported in this paper.

\clearpage

\begin{deluxetable}{cccrccl}
\tablecaption{VLA Array and Spectrometer setup for 4765 MHz Observations}
\label{tab:vlasum}
\tablewidth{
6.7in}
\tablehead{
\colhead{Date} &    \colhead{Array} & \colhead{Beam}  & \colhead{n$_{chan}$} & \colhead{$\Delta$V} & \colhead{V$_{total}$}  & \colhead{Source list} \\
   & & (\arcsec) & &(km s$^{-1}$) & (km s$^{-1}$) & }
\startdata
2000 Jan 30 & B & 2x1 & 127 &  0.384  &  47 & DR21EX  \\
2001 Mar 02 & B  & 3x1 & 127 & 0.768 & 94 & DR21EX, W75N, W3(OH), W3(C) \\
2001 Jun 12 &   B & 7x2  & 127 & 0.384 & 47 & DR21EX, W3(OH), W3(C) \\
2001 Aug 10  & C & 4x4  & 127 & 0.384 & 47 & W49N, K3-50 \\
2001 Aug 26 & C & 5x4  & 63 & 0.192 & 11 & W3(OH) \\
2001 Aug 27 & C &  9x4 & 63 & 0.192 & 11 & DR21EX\\
2001 Nov 15 & D &  14x12 & 63 & 0.192 & 11 & DR21EX, K3-50, W3(OH), W3(C) \\
2002 Jan 11 & D-A & 0.4x0.3 & 127 & 0.096 & 11& W3(OH), W3(C)  \\
2002 Apr 22 & A  &  0.4x0.3  & 127 & 0.096 & 11 &W3(OH) \\
2002 Jun 01 & BnA &   2x0.5  &127 & 0.768 & 94 & W49N \\
2002 Jul 24 & B & 1x1 & 63 & 0.192 & 11 & DR21EX, K3-50 \\
\enddata
\end{deluxetable}

\clearpage

On 2001 August 10, we observed two features in W49 with the VLA.  The peak
flux densities were 162 mJy at V$_{LSR}$= 11.89 km s$^{-1}$ and 437 mJy at
V$_{LSR}$=2.28 km$^{-1}$.  Without a phase-reference source, a flux density
$\geq$ 1 -- 2 Jy/beam is necessary to calibrate in the narrow frequency
channels used for the VLBA observations.  Because there was no
phase-reference source available for W49, detection of the signals measured
on 2001 August 10 with the VLA would have been impossible with the VLBA.

We also searched for W75N which had been observed to be highly variable in
the last 20 years (summarized by \citealt{pwg03}).  On 2001 March 02, the
peak flux density was only 33 mJy; it is not surprising that we were
unable to detect it with the VLBA after phase-referencing to DR21EX.

As a possible future VLBA target, we also searched for emission from the W3
continuum source (W3(C)).  Highly variable emission from this position has
been observed from time to time over the years \citep{gwp1,sm97}.  On 2001
June 12, we detected a 9-$\sigma$ signal (flux density: 58 mJy/beam) in a
single 0.348 km s$^{-1}$ spectral channel at $\alpha (B1950)$ = 02 21
53.29$\pm$0.06, $\delta(B1950)$ = 61 52 23.4$\pm$0.3, V$_{LSR}$=-37.6 km
s$^{-1}$ .  On 2001 March 02, we had detected a 5-$\sigma$ signal with flux
density about 50\% of the value on 2001 June 12 in a single 0.768 km s$^{-1}$
spectral channel at the same position and velocity.  These results are
consistent with non time-variable emission because of the differing
spectral resolution.

\clearpage

\begin{table}
\caption{4765 MHz Fluxes and Velocity Centroids from VLA and VLBA Observations}
\label{tab:vlavlba}
\begin{tabular}{lcccccc} \tableline \tableline 
 Date & \multicolumn{2}{c}{W3(OH)} & \multicolumn{2}{c}{DR21EX} &
 \multicolumn{2}{c}{K3-50} \\ & Flux & Velocity & Flux & Velocity & Flux &
 Velocity \\ & (Jy*km s$^{-1}$) & (km s$^{-1}$) & (Jy*km s$^{-1}$) & (km s$^{-1}$)
 & (Jy*km s$^{-1}$) & (km s$^{-1}$) \\ \hline 2000 Jan 30 & --- & --- & 2.4 &
 4.23 & --- & --- \\ 2000 Dec 23 & 1.1 & -43.94 & --- & --- & --- & --- \\
 2001 Feb 19/24 & --- & --- & 1.5 & 3.97 & 0.86 & -20.35 \\ 2001 Mar 02 &
 3.2 & -44.41 & 2.7 & 4.00 & --- & --- \\ 2001 Jun 12 & 3.4 & -44.41 &
 2.6 & 3.99 & --- & --- \\ 2001 Aug 10 & --- & --- & --- & --- & 0.84 &
 -20.26 \\ 2001 Aug 26 & 3.4 & -44.31 & --- & --- & --- & --- \\ 2001 Aug
 27 & --- & --- & 2.6 & 3.88 & --- & --- \\ 2001 Nov 15 & 3.9 & -44.26 &
 3.0 & 3.85 & 1.1 & -20.25 \\ 2002 Jan 11 & 3.8 & -44.26 & --- & --- &
 --- & --- \\ 2002 Apr 22 & 3.5 & -44.31 & --- & --- & --- & --- \\ 2002
 Jul 24 & --- & --- & 2.7 & 3.69 & 0.83 & -20.23 \\ \tableline
\end{tabular}
\end{table}

\clearpage

\section{VLBA Observational Results}
The 4765 MHz VLBA observations of W3(OH), K3-50, and DR21EX were imaged in
all four Stokes parameters.  W49 and W75N proved too weak to detect with
the VLBA.  The results from imaging are discussed in this section.  Only
preliminary results for the 1720 MHz observations in W3(OH) will be
presented here.

\subsection{W3(OH)}
\subsubsection{4765 MHz Observations}
\begin{figure}
\plotone{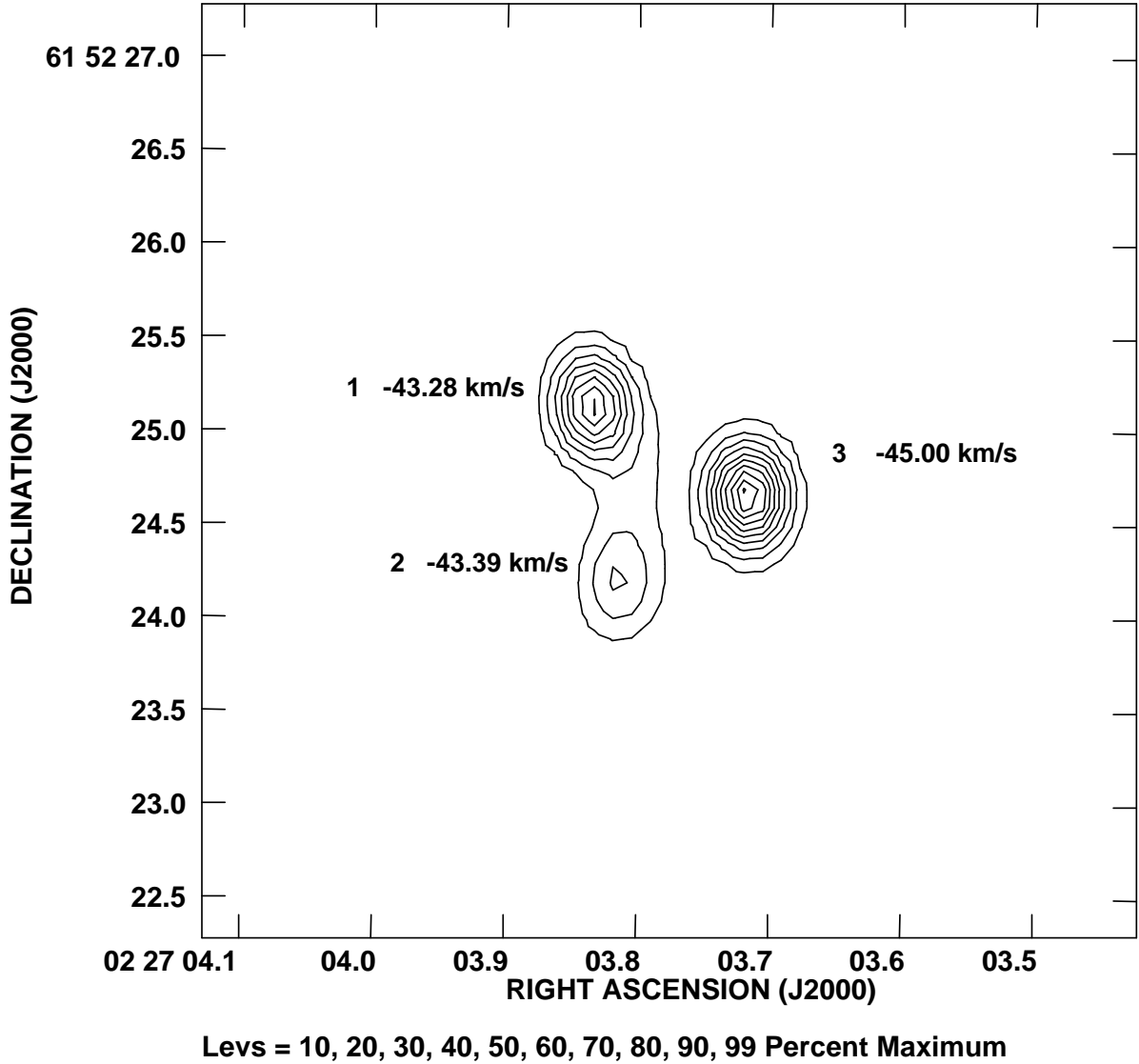}
\caption{W3(OH) integrated line emission observed with the VLA (beam: 0.4\arcsec \ x
0.3\arcsec, PA= 6\arcdeg).  The contours show $\int S(v) dv$ relative to
the largest value (feature 3).  The number used to identify the field and
the $V_{LSR}$ of the maximum in that field are shown.  Fields 1, 2, and 3
imaged with the VLBA (Figures 2, 3, and 4) correspond to the three peaks in
this VLA image.}
\label{fig:vla-a}
\end{figure}

\begin{figure}
\centering
\begin{minipage}[c]{\textwidth}
   \centering
\includegraphics[scale=.8, angle=-90, origin=5cm 3cm]{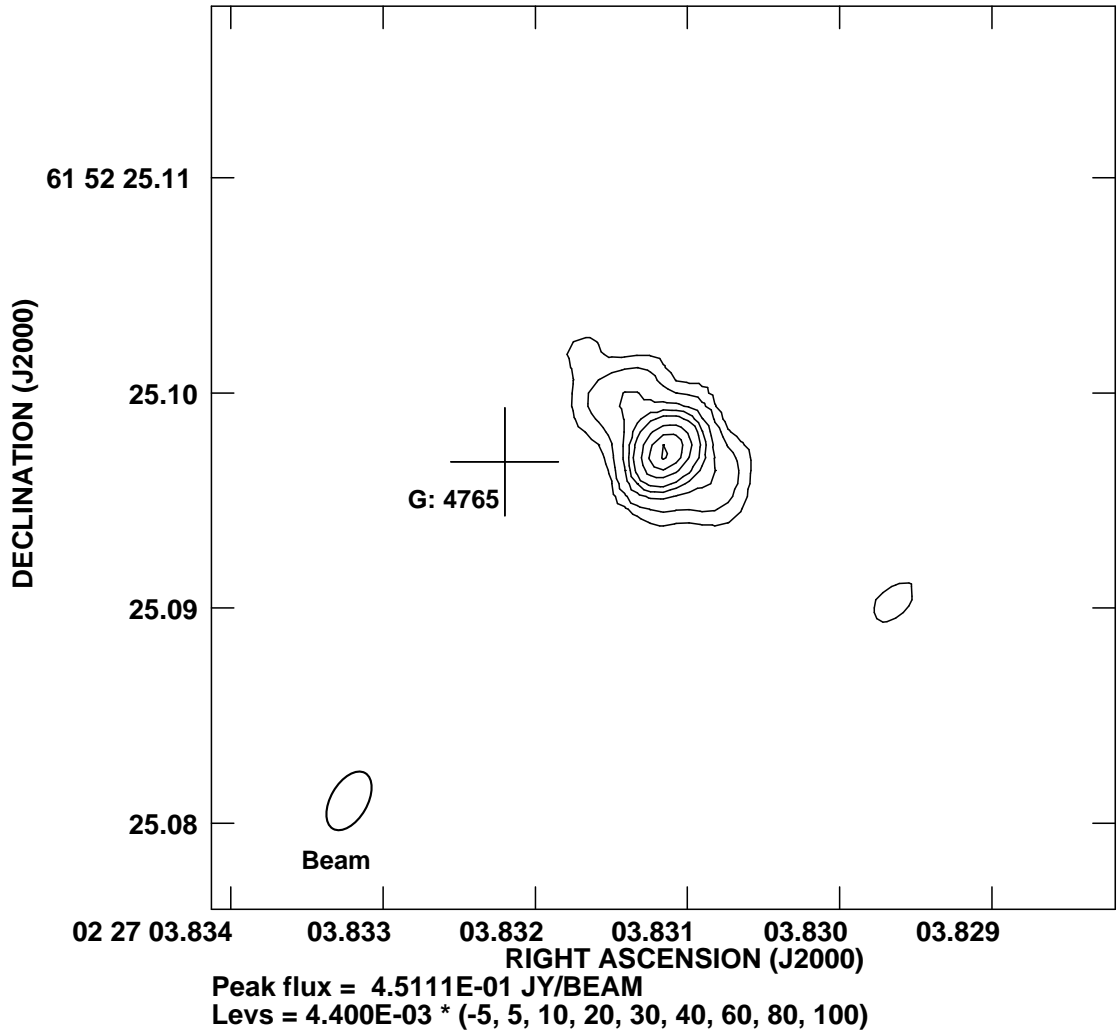}
\end{minipage}
\caption{The contours indicate the 4765 MHz emission from Field 1 of W3(OH)  
in the channel with maximum intensity (V$_{LSR}$= -43.38 km s$^{-1}$).  The
position found by GCRYF for 4765 MHz emission is indicated.  The offset
between the GCRYF position and the position determined in this paper is 8
mas.  The estimated absolute positions errors are $\pm$5 mas for GCRYF and
$\pm$1 mas for this paper.  Therefore the offset is marginally significant.
The nearest 1720 MHz features are $\sim$0.2\arcsec \ north.  Contours are
in units of the RMS noise (4.4 mJy).  The synthesized beam (2.9 mas x 1.6
mas) is shown in the lower left corner.}
\label{fig:w3fld1}
\end{figure}

\begin{figure}
\begin{minipage}[c]{\textwidth}
   \centering
\includegraphics[scale=.8, angle=-90, origin=5cm 3cm]{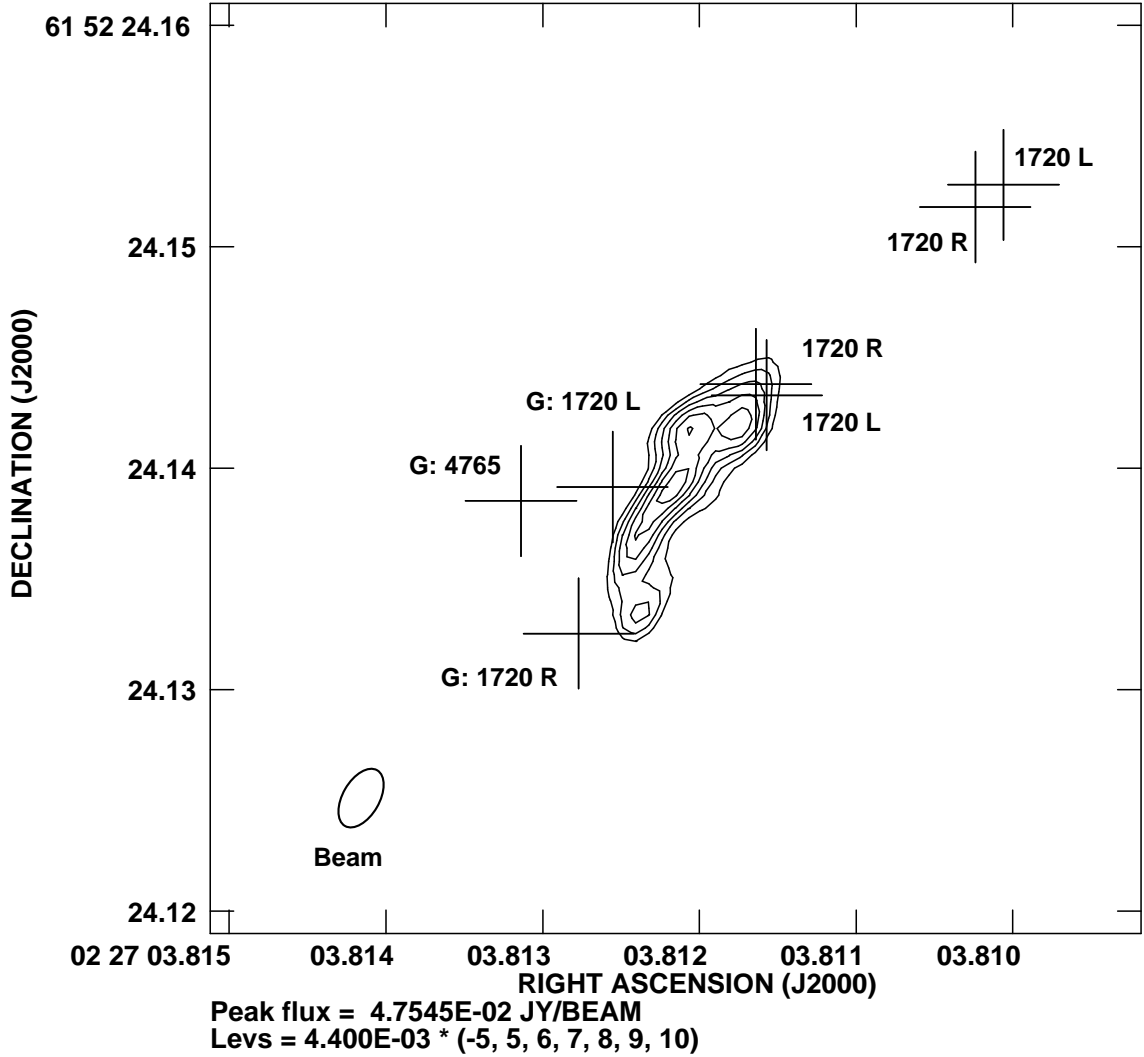}
\end{minipage}
 \caption{The contours indicate the 4765 MHz emission from Field 2 of
 W3(OH) in the channel with maximum intensity (V$_{LSR}$= -43.52 km
 s$^{-1}$).  The 1720 MHz positions are indicated by crosses.  The GCRYF
 positions for 4765 MHz and 1720 MHz emission are also indicated (prefixed
 by G:).  At 1720 MHz, the estimated absolute positions errors are $\pm$15
 mas for GCRYF and $\pm$5 mas for this paper.  Therefore the offset between
 the two determinations of 1720 MHz positions is not significant.  The
 offset between the two 4765 MHz positions is 8 mas.  At 4765 MHz, the
 estimated absolute positions errors are $\pm$5 mas for GCRYF and $\pm$1
 mas for this paper.  Therefore the offset at 4765 MHz is marginally
 significant.  Contour scheme and synthesized beam: as in Figure 2.}
\label{fig:w3fld2}
\end{figure}

\begin{figure}
 \plotone{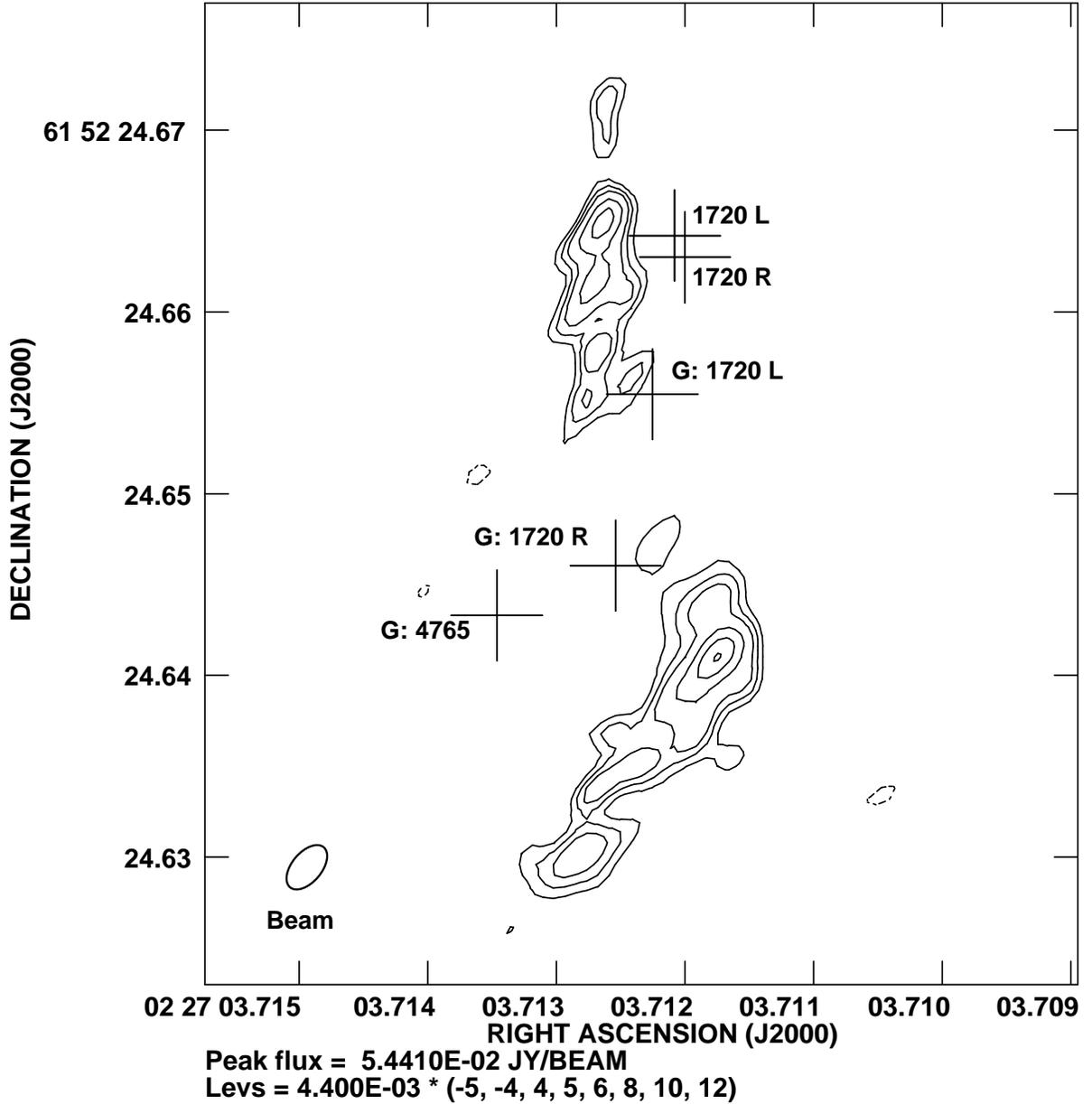} \caption{The contours indicate the 4765 MHz emission
 from Field 3 of W3(OH) in the channel with maximum intensity (V$_{LSR}$=
 -45.00 km s$^{-1}$).  The 1720 MHz positions are indicated by crosses.
 The GCRYF positions for 4765 MHz and 1720 MHz emission are indicated
 (prefixed by G:).  Note that with the increased angular resolution of our
 observations, the source is heavily resolved and at least two maxima are
 evident.  The 1720 MHz emission, although elongated N-S, does not extend
 to the location of the southern 4765 MHz complex.  Because of the absolute
 position errors in GCRYF and this paper as well as the complex source
 distribution at the current resolution, the offsets between GCRYF and this
 paper are not significant.  Contour scheme and synthesized beam: as in
 Figure 2.}
\label{fig:w3fld3}
\end{figure}

We found three regions of maser activity in the field of view. The relative
positions of the three fields (referred to as fields 1, 2 and 3) are
displayed in Figure~\ref{fig:vla-a}, which shows the emission observed with
the VLA A-array on 2002 April 22. The best estimate for the distance to
W3(OH) is 2 kpc (\citealt{gg76}; \citealt{iks00}).  Therefore, the three
fields fall in a $\sim$2000 AU diameter region.  Figures~\ref{fig:w3fld1},
~\ref{fig:w3fld2}, and ~\ref{fig:w3fld3} show these masers in greater
detail.  The positions, peak flux densities (S$_{max}$), and the
deconvolved angular sizes were obtained from Gaussian fits.  In
field 3, three Gaussian components were used, and these are labeled North,
Center, and South (N, C and S).  Data from the fits are shown in
Table~\ref{tab:vlbadata}.  The absolute accuracy of the positions is
estimated to be $\pm$1 mas while the accuracy of the relative positions
within W3(OH) is $\sim$0.1 mas.  Therefore, to within the estimated
uncertainties in absolute positions ($\pm$1 mas and $\pm$5 mas), 
the positions marginally agree with those determined by GCRYF.

\begin{figure*}
\centering
\begin{minipage}[c]{.45\textwidth}
   \centering
   \includegraphics[width=\textwidth]{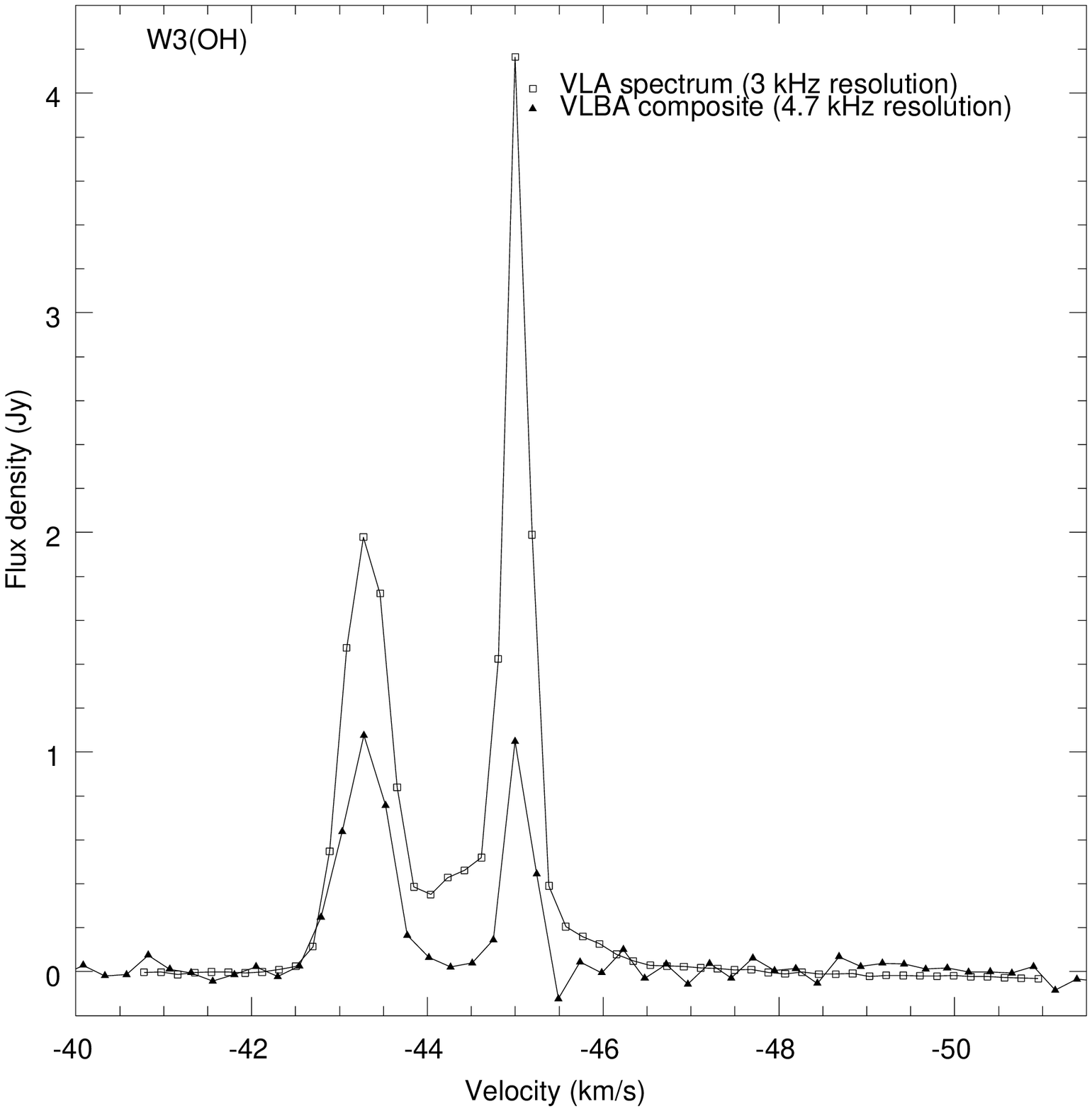}
\end{minipage}
\hspace{0.05\textwidth}
\begin{minipage}[c]{.45\textwidth}
   \centering
   \includegraphics[width=\textwidth]{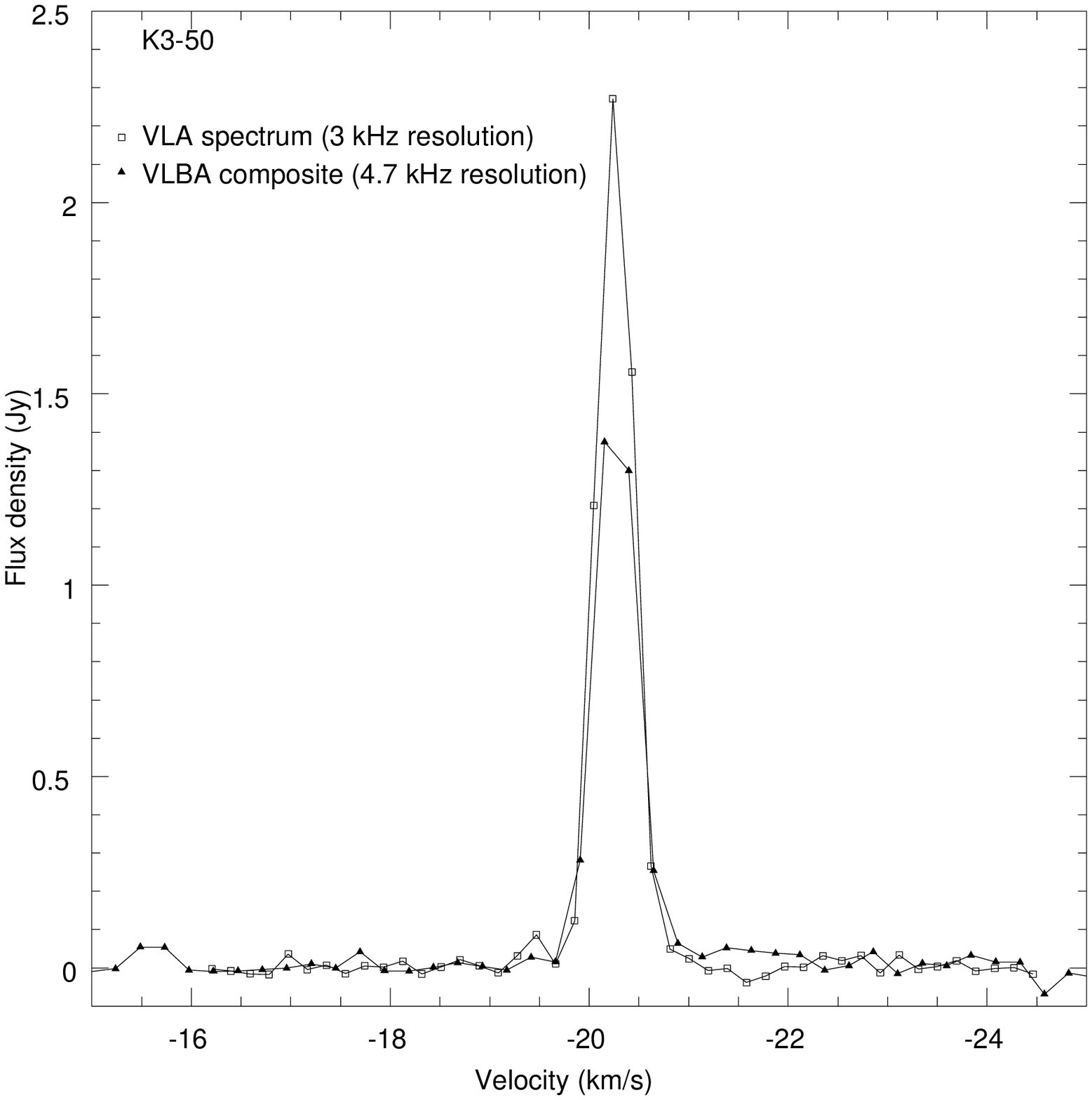}
\end{minipage} \\[4truemm]

\begin{minipage}[c]{.45\textwidth}
   \centering
   \includegraphics[width=\textwidth]{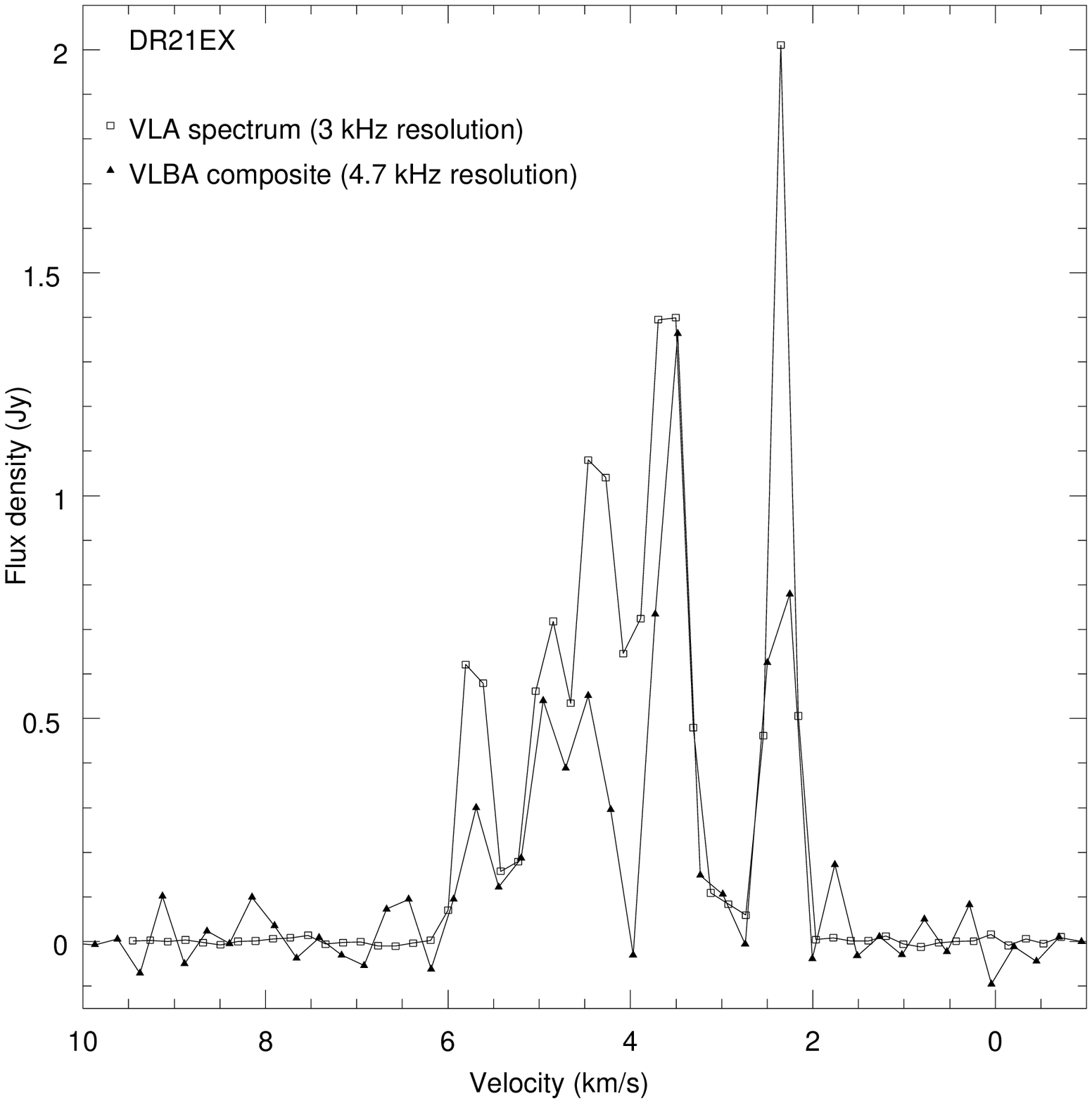}
\end{minipage}
\caption{Comparison of VLA and VLBA spectra for W3(OH), K3-50, and DR21EX.
Note that the velocity resolutions are slightly different, so that very
narrow features are expected to have somewhat lower peak flux densities in
VLBA than in VLA spectra.  The integral over the spectrum is unaffected.
See Tables 2 and 3 for observational details.}
\label{fig:vlavlba}
\end{figure*}

The minor axes of the masers are in the range 1.2 -- 5.2 mas (2.4 -- 6.2
AU).  In fields 2 and 3 the sources are elongated by factors of 3 -- 5.
Furthermore, inspection of Table~\ref{tab:vlavlba} shows that the VLBA
recovered only about 30\% of the total 4765 MHz line flux from W3(OH).
Figure~\ref{fig:vlavlba} is a superposition of the VLA and VLBA spectra.
From comparison of the spectra, it is clear that 1) all of the W3(OH) maser
features are partially spatially resolved, 2) the V=-45 km s$^{-1}$ feature
(field 3) is heavily resolved because this feature is much less intense in
the VLBA data, and 3) the ``plateau'' between the two maxima in the
VLA spectra is completely resolved by the VLBA.  Even allowing for the
difference in velocity resolution, the spectra of all of the
individual features observed with the VLBA are noticeably narrower in
velocity that the corresponding features observed with the VLA.

From the VLA observations, the position of the ``plateau'' feature is:
$\alpha(J2000)$ = 02 27 03.81, $\delta(J2000)$ = 61 52 24.3, very close to
the position of the maser in field 2.  (The uncertainty in this VLA position
is estimated to be $\pm$0.2\arcsec \ because of blending in the VLA data
with the stronger nearby features in fields 1 and 3).  Therefore, we can
exclude the possibility that the ``plateau'' was outside of our field of
view with the VLBA.  From the VLA observations, the size of the plateau
feature is $<$0.5\arcsec \ x $<$0.2\arcsec \ .  Non-detection of the plateau
in VLBA observations constrains its size to be $\geq$60 mas.  The two
sets of observations constrain the brightness temperature, $T_{b}$, of
this feature: 1.2 10$^5$\ K $\leq$ T$_b$ $\leq$ 3.2 10$^6$ K.

The polarization measurements yielded only upper limits.  For each Stokes
parameter, we integrated the flux density over the source in each field.
The upper limits for Q, U, and V are expressed as a percentage of I in
Table~\ref{tab:polar}.  From VLA observations, we an find that the upper
limit for V integrated over all of the W3(OH) masers is: V $\leq$ 0.8\% I
(4 $\sigma$).
\subsubsection{1720 MHz Observations}

The positions, peak flux densities, deconvolved angular sizes, and
V$_{LSR}$ of the peaks of the 1720 MHz features are presented in
Table~\ref{tab:1720}.  The synthesized beam is 6.2 mas x 5.5 mas at a
position angle of 49\arcdeg.  The values for both the left and right
circularly polarized (LCP and RCP) features are listed for the same fields
used to describe the 4765 MHz observations above.  The LCP and RCP features
appear to be Zeeman pairs because they have the same positions to within 2
mas.  Table~\ref{tab:4765vs1720} is a comparison between the 1720 MHz and
4765 MHz results.  In this table, the first two columns provide the
separation of the 1720 MHz LCP feature(s) from the 4765 MHz feature; the
third and fourth columns, the corresponding separations for the 1720 RCP
feature(s); the fifth, the parallel component of the magnetic field
(B$_\parallel$ ) computed from the velocity splitting
\citep{lo75,gray01}. The sixth column provides the ``demagnetized
velocity'' -- the velocity that the feature would have in the absence of
Zeeman splitting.  The last column provides $\delta V$, the difference
between the demagnetized velocity of the 1720 MHz features and the velocity
of the 4765 MHz features.  The demagnetized velocity is used in this
comparison because the 4765 MHz features have no Zeeman splitting.  The
offsets are given in the sense 1720-4765.  The sign of B$_\parallel$ is the
same at all six positions in W3(OH); values are 5.4 -- 10.0 mG.
 
In Field 1, the 1720 MHz features occur in three clumps. As a function of
velocity, the peak position moves approximately along a northwest to
southeast line, both within and between clumps.  The same pattern was
noticed by \citet{mfg94}.  The positions tabulated are the positions with
maximum intensity within each of the three clumps.  In Field 1, the nearest
1720 MHz feature is displaced $>$190 mas from the 4765 MHz position.  Near
the position of the 4765 MHz feature (see Figure~\ref{fig:w3fld1} and
Table~\ref{tab:vlbadata}), the upper limit for any 1720 MHz emission is 75
mJy (4$\sigma$).  The first feature in Field 2 in Table~\ref{tab:1720} differs in
position from the 4765 MHz feature by 5$\pm$5 mas and in velocity by
-0.06$\pm$.08 km s$^{-1}$.  The second feature differs from the 4765 MHz
feature in position by $\sim$18$\pm$5 mas and its velocity differs by
$\sim$0.6$\pm$.08 km s$^{-1}$.  In Field 3, the only 1720 MHz feature is
very elongated N-S, and differs in position with the 4765 MHz feature
called Field 3N by $\sim$5$\pm$7 mas and in velocity by 0.04$\pm$.08 km
s$^{-1}$.  While \citet{mfg94} found that some of the 1720 MHz masers had
sizes $\leq$1.2 mas, we find no features with sizes $<\sim$2 mas.  The
sizes of the 1720 MHz features in this data are all $\geq$ 2.7 mas.

Our 1720 MHz positions and those found by GCRYF with a 130 mas synthesized beam
are plotted in Figure~\ref{fig:w3fld2} and Figure~\ref{fig:w3fld3}.  In
GCRYF's study, conducted with a 40 mas beam at 4765 MHz, two of the 4765
MHz sources (Fields 2 and 3) showed spatial coincidence with 1720 MHz
emission.  This finding is supported and refined in our study with a factor
of 10 smaller beam.  In Fields 2 and 3, the positional differences are less
than our estimated absolute positional accuracy, and the velocity
differences are much less than our velocity resolution at 4765 MHz.
Therefore, to within the accuracy of our absolute position determinations
($\sim$5 mas), there are 1720 MHz masers at the same positions as 4765 MHz
masers in Fields 2 and 3N.  Nevertheless, there are no corresponding 1720
MHz masers for the 4765 MHz masers in Fields 1, 3C and 3S.  In addition,
four 1720 MHz Zeeman pairs have no corresponding 4765 MHz masers.  Further
details of the 1720 MHz observations will be reported separately.

\subsection{K3-50}
\begin{figure}
\centering
\begin{minipage}[c]{\textwidth}
    \centering
\includegraphics[scale=.9, angle=-90, origin=5cm 3cm]{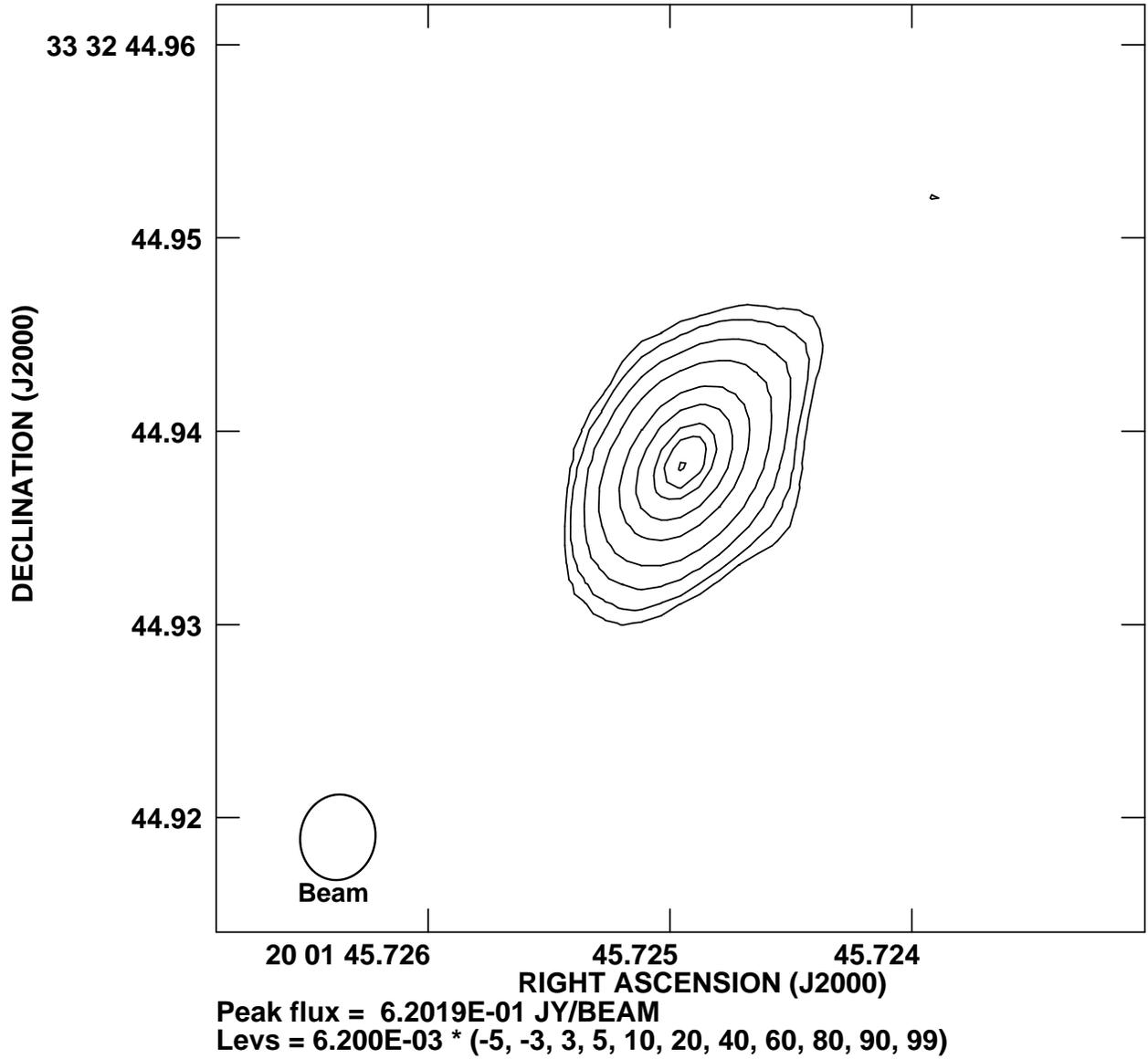}
\end{minipage}
\caption{The contours indicate the 4765 MHz emission brightness in 
the channel with maximum intensity (V$_{LSR}$= -20.15 km s$^{-1}$) toward
K3-50.  Contours are in units of the RMS noise (6.3 mJy); the synthesized
beam (4.4 mas x 3.9 mas) is shown in the lower left corner.}
\label{fig:k3-50}

\end{figure}

With the 4.4 mas x 3.9 mas beam, K3-50 is a single, slightly resolved
source. The peak flux density is $\sim$0.6 Jy/beam.  Both a comparison between
the spectra measured with the VLBA and the VLA (Figure~\ref{fig:vlavlba})
and the integrated line flux (Table~\ref{tab:vlavlba}) show
that the VLBA recovers all of the 4765 MHz line flux from this source.  The
absolute position of the maser is given in Table~\ref{tab:vlbadata}.  The
accuracy of this position is estimated to be $\pm$1 mas (see Section 2).  The
emission in the channel with maximum intensity is shown in
Figure~\ref{fig:k3-50}.  Unlike W3(OH), there is no known nearby 1720 MHz
maser.  The nearest 1720 MHz maser is ON-3, $\sim$2.3\arcmin \
northeast of the 4765 MHz maser.

The accepted distance for K3-50 is 7.4 kpc (i.e. \citet{h75}; scaled to 8.5
kpc distance to the Galactic center).  Because of the greater distance, it
is not surprising that this source is less resolved than either W3(OH) or
DR21EX.  However, like W3(OH), the source is resolved with a linear size of
6.3 mas x 2.8 mas ($\sim$ 50 x 20 AU).  The polarization upper limits are
summarized in Table~\ref{tab:polar}.

\subsection{DR21EX}
\begin{figure}
\centering
\begin{minipage}[c]{\textwidth}
   \centering
\includegraphics[scale=.8, origin=5cm 3cm]{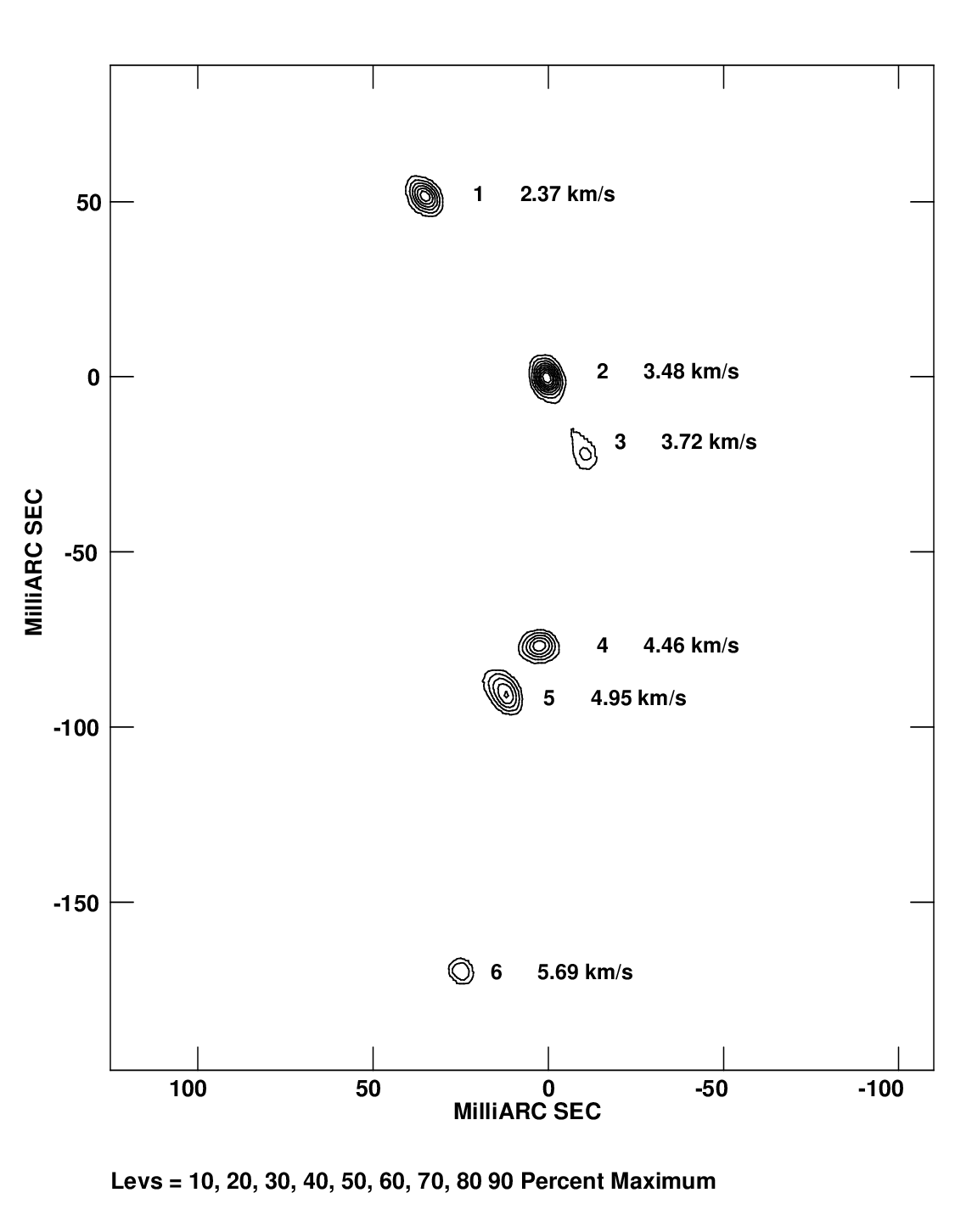}
\end{minipage}
\caption{The contours indicate positions and integrated
intensities ($\int S(v) dv$) of the six maser spots in DR21EX relative to
the position and integrated intensity of feature 2.  The number used to
identify the feature and its $V_{LSR}$ are shown.  While the absolute
positions of the features are not determined in the VLBA observation, the
error in relative positions is $\leq$.1 mas.  The synthesized beam is 5.7
mas x 4.4 mas, P.A.= 23\arcdeg.}
\label{fig:dr21ex_loc}
\end{figure}

DR21EX has six distinct maser spots which lie along an arc with angular
extent $\sim$0.25\arcsec.  At the estimated distance (2 kpc:
\citet{ddw78}), this extent corresponds to $\sim$500 AU.  The synthesized 
beam is 5.7 x 4.4 mas at a position angle of 23\arcdeg.  As explained in
Section 2, absolute positions cannot be obtained for this source due to the
lack of a phase-referencing source.  All positions were referenced to the
strongest component (feature 2).  Our best estimate for this position from
VLA observations is: $\alpha(J2000)$ = 20 39 00.377$\pm$.008,
$\delta(J2000)$ = 42 24 37.28$\pm$.07.  The relative locations of the six
spots are shown in Figure~\ref{fig:dr21ex_loc}.  The six spots are
approximately along a N-S arc, and their velocities increase from North to
South.  The measured parameters for each spot are presented in
Table~\ref{tab:vlbadata}.  The positions (referenced to the VLA position of
feature 2), peak flux densities, LSR velocities, and deconvolved sizes are
summarized in Table~\ref{tab:vlbadata}. The degree of resolution varies
significantly from feature to feature (see Figure~\ref{fig:vlavlba}).  The
minor axes of the spots range from $\sim$4 -- $\sim$8 AU, and they are
typically elongated by less than a factor of two.  Comparison of the flux
measured with the VLA shows that the VLBA observation recovers about 60\%
of the total flux (see Table~\ref{tab:vlavlba}).  Therefore, a significant
component of 4765 MHz emission is not detected by the VLBA.  As was true
for W3(OH), the components detected with the VLBA have noticeably narrower
velocity widths that those detected with the VLA.

In VLA B-array observations on 2000 January 30, several 1720 MHz OH maser
features were found within 0.2\arcsec \ of the 4765 MHz masers in this
source (P. Palmer \& W. M. Goss, in preparation).  However, improved angular resolution at 1720 MHz and
phase-referencing for both frequencies are needed to determine if the 4765
MHz and 1720 MHz masers coincide.  The results of the polarization
measurements are provided in Table~\ref{tab:polar}.

\clearpage

\begin{table}
\begin{minipage}{125mm}
\caption{4765 MHz Maser Parameters from VLBA Observations}
\label{tab:vlbadata}
\begin{tabular}{llccccl}
\tableline \tableline
 Source & Field & $\alpha_{J2000}$&$\delta_{J2000}$& S$_{max}$ & V$_{LSR}$ & 
Size \\
  & & & & (mJy/bm) & (km s$^{-1}$) & (mas) \\
\tableline 
W3(OH) &  1   & 02 27 03.8311 & 61 52 25.097 & 410 & -43.28 & 2.6 x 0.9 \\
W3(OH) &  2   & 02 27 03.8120 & 61 52 24.140 & 42 & -43.39 & 13 x 3.9 \\
W3(OH) & 3 N &  02 27 03.7126 & 61 52 24.661 & 33 & -45.0 & 20 x 3.9 \\
W3(OH) & 3 C &  02 27 03.7118 & 61 52 24.642 & 32 & -45.0 & 11 x 4.0 \\
W3(OH) & 3 S & 02 27 03.7125 & 61 52 24.633 & 26 & -45.0 & 15 x 5.5 \\
K3-50  &     & 20 01 45.7249 & 33 32 44.938 & 620 & -20.35 & 6.3 x 2.8 \\
DR21EX & 1   & 20 39 00.3801 & 42 24 37.332 & 540 & 2.37 & 5.2 x 2.9 \\
DR21EX\footnote{All DR21EX
positions are referenced to the VLA determination of the position of this
component (see text).} & 2   & 20 39 00.3770 & 42 24 37.280 & 890 & 3.48 & 4.7 x 3.2 \\
DR21EX & 3   & 20 39 00.3759 & 42 24 37.259 & 250 & 3.72 & 5.1 x 3.9 \\
DR21EX & 4   & 20 39 00.3772 & 42 24 37.204 & 380 & 4.46 & 6.2 x 1.8 \\
DR21EX & 5   & 20 39 00.3780 & 42 24 37.190 & 440 & 4.95 & 6.8 x 3.7 \\
DR21EX & 6   & 20 39 00.3791 & 42 24 37.111 & 320 & 5.69 & 4.1 x 3.1 \\
\tableline
\end{tabular}
\end{minipage}
\end{table}

\begin{table}
\begin{minipage}{125mm}
\caption{Upper Limits for 4765 MHz Polarization}
\label{tab:polar}
\begin{tabular}{llccc}
\tableline \tableline
 Source & Field & Q & U & V \\  
        &      & (\% of I)& (\% of I)& (\% of I) \\
\tableline
W3(OH) & 1 & $\leq$2 & $\leq$2 & $\leq$4 \\
W3(OH) & 2 & $\leq$12 &$\leq$12 &$\leq$12 \\
W3(OH) & 3 & $\leq$12 &$\leq$12 &$\leq$7 \\
K3-50  &  &  $\leq$4 &$\leq$4 &$\leq$4 \\
DR21EX & 1 & $\leq$5 &$\leq$5 &$\leq$5 \\
DR21EX & 2 & $\leq$3 &$\leq$3 &$\leq$3 \\
DR21EX & 3 & $\leq$11 &$\leq$11 &$\leq$11 \\
DR21EX & 4 & $\leq$6 &$\leq$6 &$\leq$6 \\
DR21EX & 5 & $\leq$6 &$\leq$6 &$\leq$6 \\
DR21EX & 6 & $\leq$11 &$\leq$11 &$\leq$11 \\
\tableline
\end{tabular}
\end{minipage}
\end{table}

\begin{table}
\begin{minipage}{125mm}
\caption{1720 MHz VLBA Observations of W3(OH)}
\label{tab:1720}
\begin{tabular}{llccc}
\tableline \tableline
$\alpha_{J2000}$ &$\delta_{J2000}$ &
S$_{max}$ & Size  & V$_{LSR}$  \\
    &  & (Jy/bm) & (mas) &(km s$^{-1}$)  \\
\tableline
LCP & & & &   \\
Field 1 & & &  & \\ 
02 27 03.8364 & 61 52 25.317 & 9.1 & 2.9x2.5 & -43.47 \\
02 27 03.8336 &  61 52 25.305& 0.15& 4.0x1.9 & -44.91 \\
02 27 03.8324 & 61 52 25.295 & 20.8 & 3.7x2.4 & -45.60 \\
Field 2 & & & & \\ 
02 27 03.8116 & 61 52 24.144 & 2.9 &7.2x3.6 & -43.77 \\ 
02 27 03.8101& 61 52 24.153 & 0.21 &7.1x5.2 & -44.53 \\ 
Field 3 & & & & \\
02 27 03.7121 &  61 52 24.664 & 0.15 &35x7 &  -45.26 \\
\tableline 
RCP & & & & \\
Field 1  & & & & \\ 
02 27 03.8364 & 61 52 25.317& 14.6& 2.8x2.6 &  -42.70 \\
02 27 03.8338 & 61 52 25.306& 0.12& 4.5x3.3&  -43.98 \\
02 27 03.8323 & 61 52 25.293 & 15.5& 3.8x2.5& -44.87  \\
Field 2 & & & & \\ 
02 27 03.8116&  61 52 24.143& 2.7 & 7.4x3.8 & -43.13 \\
02 27 03.8102 & 61 52 24.152& .88 & 6.4x4.3 & -43.43 \\
Field 3 & & & & \\
02 27 03.7120 &  61 52 24.663 & 0.13 &31x8 &  -44.66  \\
\tableline
\end{tabular}
\end{minipage}
\end{table}

\begin{table}
\begin{minipage}{125mm}
\caption{Comparison of 4765 and 1720 MHz Positions and Velocities in W3(OH)}
\label{tab:4765vs1720}
\begin{tabular}{ccccccc}
\tableline \tableline
$\Delta \alpha$  & $\Delta \delta$ & $\Delta \alpha$  & $\Delta \delta$ & B$_{\parallel}$ & V$_{*}$ & $\delta$V \\ 
 (mas) & (mas) &  (mas) & (mas) &  (mG) & (km s$^{-1}$) & (km s$^{-1}$) \\
\tableline
LCP  &       &  RCP  &      &    &        &      \\
Field 1 & & & & & & \\
38 & 220 & 37.1  & 220 & 7.0  & -43.08 & 0.20 \\
18 &  208 & 20 & 209 & 8.4  & -44.44 & -1.63 \\
 9.5&  198  &  8.3 & 196.4 & 6.6  & -45.24 & -1.96 \\
Field 2 & & & & & & \\
-2.5 &  3.8 & -3.0  & 3.3 & 5.8 & -43.45 &  -0.06 \\
 -14 & 13 & -12 &12 & 10.0 & -43.98 &  -0.59 \\
Field 3\footnote{Comparisons with 4765 MHz Field 3N} & & & & & & \\
 -3.7 & 3.2 &  -4.2 & 2.0 & 5.4 & -44.96 & +0.04 \\
\tableline
\end{tabular}
\end{minipage}
\end{table}

\clearpage

\section{Discussion}

We provide an overview of the observational results:
\begin{itemize}
\item Co-propagation of 4765 MHz and 1720 MHz emission:

In W3(OH), the maser positions in Fields 2 and 3 are compatible with models
that require coincidence between the masers in the two transitions.
However, Field 1 shows no such coincidence.  Therefore, it is clear that
not all 1720 MHz masers have detectable 4765 MHz counterparts and not all
4765 MHz features have detectable 1720 MHz counterparts; however, in some
cases 4765 MHz and 1720 MHz emission can arise from regions separated by
$<$10 AU.  \citet{lge99} find that as population in the $^2\Pi_{1/2},
J=1/2$ \ state increases in their models, the far-IR lines connecting this
state to the ground $^2\Pi_{3/2}, J=3/2$ \ state ``turn off'' the 1720 MHz
maser.  However, a detectable 4765 MHz line requires that the $^2\Pi_{1/2},
J=1/2$ \ state is populated.  Producing the two masers simultaneously seems
likely only for a narrow range of conditions.  Therefore, the observations
are compatible with co-propagation models if these models allow widely
varying intensity ratios of 4765 MHz to 1720 MHz emission in order to
produce observable intensities in either of the masers or both.

\item Angular Sizes:

In both W3(OH) and DR21EX, a significant fraction of the flux was resolved
out with the VLBA; nevertheless a number of features have minor axes
ranging from $\sim$1 -- 5 mas.  Many of the features were elongated by
$\geq$ 3:1.  The picture that emerges is that OH 4765 MHz emission occurs
in regions 100's of AU in diameter in these sources, and that 30 -- 60\% of
the flux density is confined to sources with diameters of 10's of AU.
Furthermore, typically emission occurs over a velocity range $\sim$4 km
s$^{-1}$, however, emission from the individual small spots has narrow
velocity widths (typically $\leq$0.2 km s$^{-1}$).

\item Brightness Temperatures:

Brightness temperatures for the 4765 MHz maser positions measured with the
VLBA are summarized in Table~\ref{tab:tab5}.  The 4765 MHz maser brightness
is listed in the third column; the corresponding 1720 MHz brightness in the
fourth.  The brightness temperatures at 4765 MHz range from 2 x 10$^7$\ K
to 5 x 10$^9$\ K.  Assuming that the fraction of flux density not detected
with the VLA arises from sources $>$60 mas, the upper limit for the
brightness temperatures of the diffuse components is T$_b \leq$ 3 10$^6$\ K
(see section 4.1.1).  The 1720 MHz brightness exceeds that at 4765 MHz by
factors of 10 -- 1000.
\end{itemize}
The above summary may be used to place constraints on the physical 
conditions in the regions of maser emission.

\clearpage

\begin{table}
\caption{Source Brightness Temperatures at 4765 MHz Maser Positions}
\label{tab:tab5}
\begin{tabular}{lccc}
\tableline \tableline 
Source & Field & $T_{b}(4765)$&  $T_{b}(1720)$ \\
  & & (K) & (K) \\
\tableline 
W3(OH) & 1   & 5 10$^9$ &  \\
W3(OH) & 2   & 4 10$^7$ & 4 10$^{10}$ \\
W3(OH) & 3 N & 2 10$^7$ & 2 10$^8$  \\
W3(OH) & 3 C & 4 10$^7$  &  \\
W3(OH) & 3 S & 2 10$^7$ &   \\
K3-50 &      & 2 10$^9$ &  \\
DR21EX & 1   & 2 10$^9$ &  \\
DR21EX & 2   & 3 10$^9$ &  \\
DR21EX & 3   & 6 10$^8$ & \\
DR21EX & 4   & 2 10$^9$ &    \\
DR21EX & 5   & 9 10$^8$ &   \\
DR21EX & 6   & 1 10$^9$ & \\
\tableline
\end{tabular}
\end{table}

\clearpage

(1) The state of saturation of the masers: the rate of depopulation of the
upper level by stimulated emission is $W_{stim}$ = $A_{ul}\frac{kT_b}{h
\nu}\frac{\Omega_M}{4 \pi}$, where $\Omega_M$ is the solid angle from which
a typical molecule in the source receives radiation at brightness
temperature $T_b$.  For the 4765 MHz line, $A_{ul}$ = 3.89 10$^{-10}$\
s$^{-1}$ \citep{dmbb77}, and we take $\frac{\Omega_M}{4 \pi} \sim$1/2 to
obtain $W_{stim}$ = 9 10$^{-9} T_b$\ s$^{-1}$.  We estimate the collisional
rate as $W_{col} \sim 10^{-10} n_{H_2}$\ s$^{-1}$.  Therefore, if $T_b
\sim$ 10$^{10}$\ K, $W_{stim}$ will exceed other rates as long as n$_{H_2}
\leq 9$ 10$^{11}$\ cm$^{-3}$.  (All OH pumping models to date assume
densities $\ll$10$^{11}$\ cm$^{-3}$).  Therefore, this brightness
temperature implies that the maser is saturated unless the emission region
is significantly elongated along the line of sight so that the value of
$\Omega_M$ is seriously overestimated above.  However, for the surrounding
regions with lower brightness, the density upper limit to avoid saturation
is $n_{H_2} \leq 3 \ 10^7$\ cm$^{-3}$.  Thus, the high brightness
temperature spots are almost certainly saturated, although the more
extended 4765 MHz emission may not be.  It has frequently been assumed
that rapid time-variations are an indicator of an unsaturated maser
(e.g. \citet{ppf94,cven95,dm98}).  If this assumption is correct for the
4765 MHz masers reported in this paper, the conclusion from the above
consideration of the saturation state of the maser is in conflict with
the observational data for 4765 MHz masers.

(2)  We examine the maser optical depth required to produce a
brightness $T_b$:  the line center optical depth is:
\begin{equation}
\tau_0 = \sqrt{\frac{4 ln(2)}{\pi}} \frac{c^2}{8 \pi
\nu^2}\frac{A_{ul}}{g_u \Delta \nu}\Delta n L,
\end{equation}
where $\Delta n$ is the population difference (per magnetic sub-level) of
the upper ($u$) and lower ($l$) energy levels involved in the transition,
and L is the pathlength.
That is,
\begin{equation}
\Delta n =\frac{n_l}{g_l}-\frac{n_u}{g_u}.
\end{equation}
Inserting appropriate constants for this transition and $\Delta \nu$=3 kHz
($\sim$0.2 km s$^{-1}$, corresponding to the narrow lines observed with the
VLBA), equation(1) becomes:
\begin{equation}
\tau_0 = 6.4\ 10^{-14} \Delta n \ L.
\end{equation}
To express $\Delta n$ in terms of the total density of molecular gas, we
write:
\begin{equation}
\Delta n = -\onehalf \eta y f n_{H_2}.
\end{equation}
Here, $\eta$ is the inversion efficiency ($\frac{n_u - n_l}{n_u+n_l}$;
\citealt{lge99}), 
$y$ is the fraction of OH molecules in the $^2\Pi_{1/2}, J=1/2$ state, f is
the fractional abundance of OH compared to H$_2$, and the factor 1/2 is
present because the population of the state is approximately equally
distributed among the sub-levels of the four hyperfine levels at plausible
temperatures and inversion fractions.

We express the pathlength in AU, and obtain:
\begin{equation}
\mid\tau_0\mid = 0.48 \ y \eta f n_{H_2} L_{AU}.
\end{equation}
For temperatures between 50 -- 200 K, $y$ is in the range 0.03 -- 0.1, and
 $\eta$ is estimated to lie in the range 0.01 -- 0.1.  Taking
the geometric mean of possible values of the product, $y \eta$ = 1.3
10$^{-3}$, we obtain:
\begin{equation}
\mid\tau_0\mid = 8\ 10^{-4} \ f n_{H_2} L_{AU}.
\end{equation}
For an OH abundance fraction f$\sim 10^{-5}$ (see \citealt{pk96a} and
references therein),  we obtain:
\begin{equation}
\mid\tau_0\mid = 8 \ 10^{-9} \ n_{H_2} L_{AU}.
\end{equation}
There is little doubt that the product $y \eta$ has an uncertainty of less
than an order of magnitude.  While f is at the upper end of possible values
for relative abundance of OH to H$_2$ in the interstellar medium
(i.e. within a factor of a few of the total O abundance), all models for OH
masers to date have invoked an increased OH fractional abundance.

The brightness temperature of the background is $\leq$ 10\ K for the 4765 MHz
masers reported in this paper.  Therefore, in order to produce $T_b \sim5 \
10^{9}$\ K, a minimum gain corresponding to $\mid\tau_0\mid \sim$ 20 is
required.  While the brightness temperatures are typically a factor of 10
less than is typical for OH masers
\citep{el92}, these values of $T_b$ still lead to rather severe constraints
on parameters of the region.  From equation (7) we find:
\[ L_{AU} = \frac{2 \ 10^9}{n_{H_2}}.  \]
Therefore if $L_{AU}$ is to be less than the diameter of the entire 4765
MHz maser region (taken to be 10$^3$ AU) we require n$_{H_2} \geq 2 \ 10^6$
\ cm$^{-3}$.  On the other hand, if the pathlength were comparable to the
spot sizes ($\sim$10 AU), required densities would be increased by a factor
of $\sim$100.  We regard such a density as implausible (see also next
paragraph). The density required can be reduced by a factor of 10 if the
product $y \eta$ were taken to be at the extreme high end of the plausible
range; however, the density would be increased by a factor of 10 if the
product were at the low end.  (The required density would be increased by
another factor of 10 if the OH abundance were similar to that found in
quiescent, low density regions of molecular clouds.)

The small measured transverse extent of each feature together with the
$\sim$100 times greater pathlength along the line of sight inferred from
limits on the density leads to the conclusion that the maser region is
filamentary with the long direction pointed toward the observer.  Such a
structure also relieves the constraint on saturation derived above since
$\Omega_M$ would be greatly reduced.  These filaments need not be geometric
structures, but may simply be lines of sight through the much larger region
in which the velocity coherence is greatest.  Such an interpretation fits
naturally with the very narrow line widths detected with the VLBA.  Where
velocity coherence does not exist, lower brightness emission may be produced.

While many models for OH masers have been developed in the last 30 years,
parity assignments for $\Lambda$-doublet levels of OH were incorrect until
1984, and collision cross sections before 1994 were significantly discrepant
compared with the best current values \citep{pk96a}.  Therefore, we
primarily consider models developed since 1994.

First, we compare the density limit derived above for the 4765 MHz maser
region with that derived from consideration of the 1720 MHz line.
\cite{lge99} find that the 1720 MHz maser is quenched by collisions at
densities above $\sim$5 10$^5$cm$^{-3}$.  Therefore the density estimate
derived above (10$^{6\pm 1}$ cm$^{-3}$) can be compatible with 1720 MHz
emission arising in the same physical region as 4765 MHz emission only if
the density is near its lower limit.  \citet{pk96a,pk96b} have explicitly
discussed the production of the 4765 MHz line and find no range of
conditions in which simultaneous production of 4765 MHz and 1720 MHz masers
can occur.  In an earlier detailed model,
\cite{gray92} can obtain coincidence of the 4765 MHz and 1720 MHz region with
a density 2 10$^7$\ cm$^{-3}$ in an accelerating flow.  Apparently,
radiative pumping in the flow is sufficient to overcome quenching; then the
accelerating flow prevents the optical depth in the far IR lines connecting
the $^2\Pi_{3/2},J=3/2$ and $^2\Pi_{1/2}, J=1/2$ states from becoming
large enough to quench the 1720 MHz maser.  Because of the difficulty other
models encounter finding physical conditions in which 1720 MHz and 4765 MHz
emission both arise, further work on this model is desirable to verify
if the results are qualitatively similar based on more recent collision cross
sections.

A model using these recent collision cross sections was developed for 1720
MHz emission near supernova remnants \citep{lge99}. This model also
explains the enhanced OH abundance in a natural way.  In this model, OH is
formed behind a C-shock surrounding the supernova remnant.  However, 4765
MHz emission is not explicitly discussed.  In regions in which 4765 MHz
emission is detected, there are certainly shocks caused by molecular
outflows.  However, these shocks are expected to be J-shocks.  If a
transverse magnetic field is present, an initial J-shock evolves into a
C-shock (see e.g. \citealt{fetal03}).  The 5 -- 10 mG magnetic fields found
in the 1720 MHz observations are suggestive of C-shocks (see
Table~\ref{tab:4765vs1720}).

A requirement for future progress is the development of models for the 4765
MHz masers similar to the \citet{lge99} models for 1720 MHz masers.  These
new models must use the improved collision cross sections and treat the
transfer of the infrared lines which connect 1720 MHz and 4765 MHz lines.

In W3(OH) we confirmed with higher positional accuracy ($\leq$ 5 mas) the
agreement between the 4765 MHz and 1720 MHz maser positions.  We also
showed that to within VLA accuracy ($\sim$0.2\arcsec), there are 1720 MHz
masers in the region of 4765 MHz emission in DR21EX.  There are no known
1720 MHz features in the Field 1 of W3(OH) and the K3-50 sources.  We also
detected no polarization for any of the 4765 MHz features.  Upper limits in
the best cases are $\sim$3\% (4$\sigma$).

VLA observations of K3-50 at 1720 MHz should be done to confirm
the absence of a 1720 MHz maser at that location.  Also, it is
important to determine absolute positions for both 4765 MHz emission and
1720 MHz emission from DR21EX.

\acknowledgments
K. Devine would like to thank the NRAO and the NSF for making this research
experience possible and C. Brogan and E. Greisen for their assistance
throughout the summer.  P. Palmer thanks the NRAO for hospitality during 
several extensive visits while this work was carried out.


\clearpage


\end{document}